\documentclass[pdflatex,iicol,sn-mathphys-num]{sn-jnl}% Math and Physical Sciences Numbered Reference Style 
\usepackage{graphicx}%
\usepackage{multirow}%
\usepackage{amsmath,amssymb,amsfonts}%
\usepackage{amsthm}%
\usepackage{mathrsfs}%
\usepackage[title]{appendix}%
\usepackage{xcolor}%
\usepackage{textcomp}%
\usepackage{manyfoot}%
\usepackage{booktabs}%
\usepackage{algorithm}%
\usepackage{algorithmicx}%
\usepackage{algpseudocode}%
\usepackage{listings}%
\usepackage{subcaption}%subfigs
\usepackage{float}
\usepackage{placeins}% FloatBarrier
\usepackage{orcidlink}
\newcommand{\un}[1]{~\mathrm{#1}}
\theoremstyle{thmstyleone}%
\theoremstyle{thmstyletwo}%

\theoremstyle{thmstylethree}%

\begin{document}

\title[Performance of Silicon photomultipliers at low temperatures]{Performance of Silicon photomultipliers at low temperatures}


\author*[1]{\fnm{Otto} \sur{Hanski} \orcidlink{0000-0002-3230-2332}}\email{otolha@utu.fi}
\equalcont{These authors contributed equally to this work.}

\author[1]{\fnm{Tom} \sur{Kiilerich} \orcidlink{0009-0004-9322-8633}}\email{tckiil@utu.fi}
\equalcont{These authors contributed equally to this work.}

\author[1]{\fnm{Sampsa} \sur{Ahopelto}}\email{srahop@utu.fi}

\author[1]{\fnm{Aleksei} \sur{Semakin} \orcidlink{0000-0001-7690-2076}}\email{assema@utu.fi}

\author[1]{\fnm{Janne} \sur{Ahokas} \orcidlink{0000-0001-5857-1541}}\email{jmiaho@utu.fi}

\author[1]{\fnm{Viacheslav} \sur{Dvornichenko}}\email{viacheslav.dvornichenko@utu.fi}

\author[1]{\fnm{Sergey} \sur{Vasiliev} \orcidlink{0000-0002-5949-1276}}\email{servas@utu.fi}

\affil[1]{\orgdiv{Department of Physics and Astronomy}, \orgname{ University of Turku}, \orgaddress{\street{Vesilinnantie 5}, \city{Turku}, \postcode{20014}, \country{Finland}}}


\abstract{We present experimental results of characterization of Silicon photomultipliers (SiPM) in a temperature range from 90~mK to 40~K. Two SiPMs, one from ONSEMI and one from Hamamatsu Photonics were tested. Operating voltage ranges, dark count rates, afterpulsing effects and photon detection efficiencies (PDE) were determined with illumination by 275 and 470~nm light fed into the cryostat via an optical fiber. A cryogenic shutter provided a true dark condition, where thermal radiation from room temperature is shielded and the thermal excitations in the chips are frozen. A second tunneling breakdown was observed at this condition, which substantially limits the operating voltage range for the temperatures 20-30 K. Below $\sim$5 K, both SiPMs recover to an operating over-voltage range of 3-5 V.  We found the chips function through the entire tested temperature range, and are capable of withstanding thermal cycling with no major performance degradation.}




\maketitle

\section{Introduction}\label{sec:intro}

A multitude of scientific experiments \cite{UTUmagtrap,soter2014segmented,OnSemi4K,LAscintillation} require cryogenic single photon detectors. The traditional choice for a single photon detector is a photomultiplier tube, but in cryogenic conditions PMTs can be difficult to work with due to their typically large form factors and high operating voltages. 

In the interest of finding a more suitable detector, we have characterized the operational parameters and photon detection efficiencies of two commercially available silicon photomultipliers from room temperature to the sub-Kelvin regime. These SiPMs are single photon avalanche diode (SPAD) based multipixel photon counter (MPPC) devices.  

Our motivation for studying SiPMs at low temperatures is to measure their characteristics as detectors for our project in two-photon spectroscopy of the 1S-2S transition in ultracold, magnetically trapped atomic hydrogen \cite{UTUmagtrap}. The experiment requires detection of extremely low fluxes of fluorescent Lyman-alpha radiation. SiPMs are an attractive alternative to cryogenic PMTs in this experiment due to their relatively low operating voltages in the range of tens of volts, along with high PDEs and small form factors.

\section{Measurement setup}\label{sec:MeasSetup}

We tested and compared two commercially available SiPM models, Hamamatsu S13370-6050CN~\cite{hamamatsudatasheet} and Onsemi MicroFJ-30035-TSV~\cite{onsemidatasheet}. They were tested at two wavelengths, visible blue at $470\un{nm}$ and UV at $270\un{nm}$. The Hamamatsu chip was chosen because it is specifically designed for UV detection, with a room temperature photon detection efficiency (PDE) of $17.5~\%$ at $270\un{nm}$ and $12~\%$ at $121\un{nm}$. The chip has also previously been tested at cryogenic and sub-Kelvin temperatures \cite{Zhang_SiPMs,HamamatsuCryo2}. The Onsemi chip was selected due to also having been tested at cryogenic temperatures \cite{OnSemi4K} and its significantly lower operational voltage compared to the Hamamatsu chip. Additionally it is manufactured at a significantly lower price point, making it an attractive alternative especially for larger detector arrays. 

A notable difference in the Hamamatsu chips compared to Onsemi is the use of metallic thin film quench resistors for the SPAD microcells. This helps reduce the effect of temperature on pulse shape, as the metal resistor resistance does not significantly change on cooling into cryogenic region. The Onsemi MicroFJ chip uses semiconductor quench resistors. Additionally, the Onsemi chip has a built-in fast output mode, which uses a high pass filter to filter out the typical slow decaying tail of an avalanche voltage pulse. For the Hamamatsu chip there is no built-in fast signal option, so we instead used a self-made highpass RC filter with a cutoff at $31.8\un{MHz}$ ($R = 50\un{\Omega}$, $C = 100 \un{pF}$).

\begin{figure}[h]
    \centering
    \begin{subfigure}{\linewidth}
    \centering
        \includegraphics[width=0.8\linewidth]{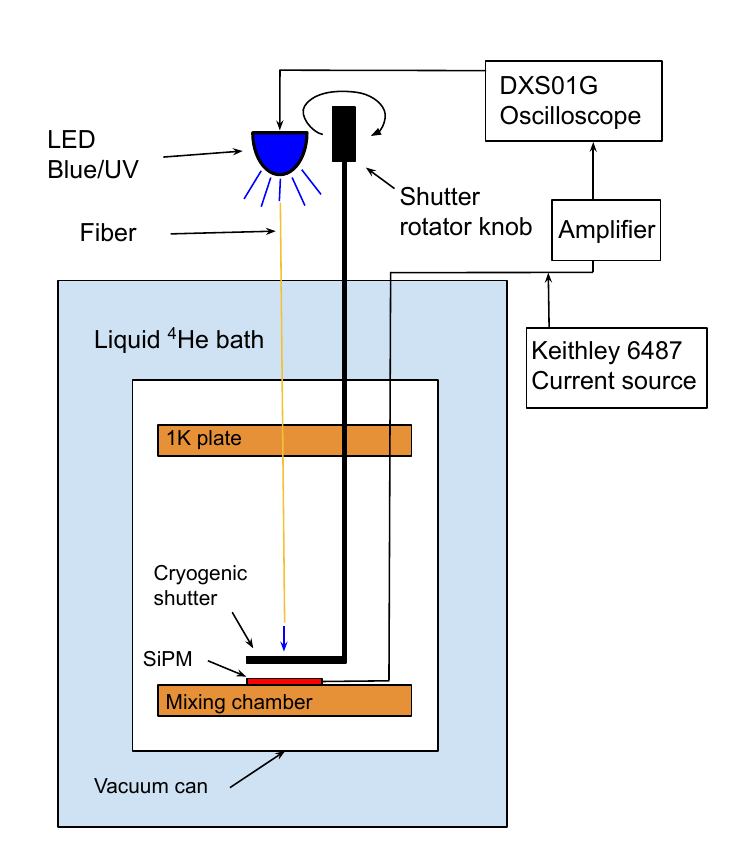}
        \caption{The measurement scheme.}
        \label{fig:measscheme}
    \end{subfigure}
    \\
    \begin{subfigure}{\linewidth}
    \centering
        \includegraphics[width=0.8\linewidth]{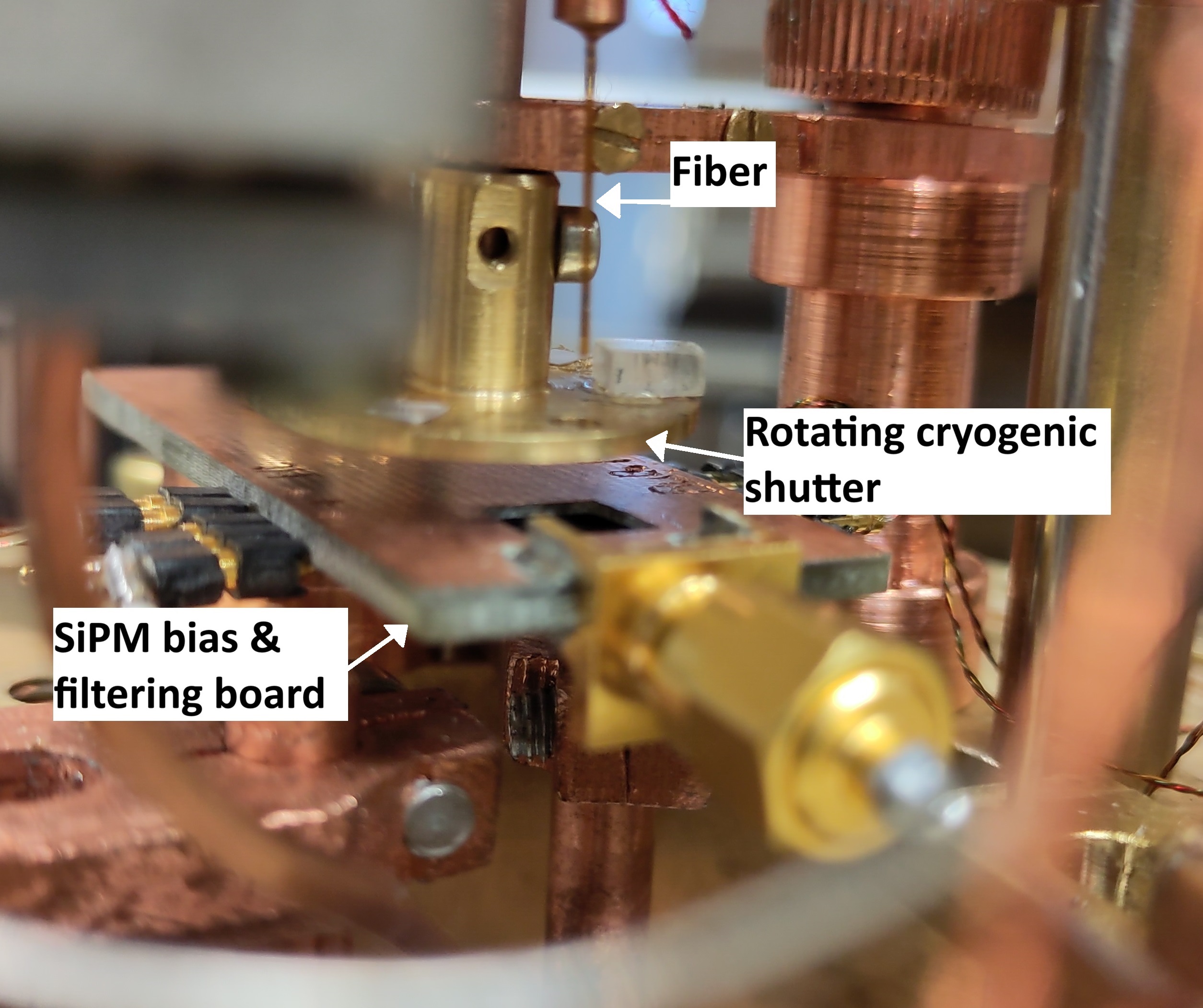}
        \caption{SiPM installation at mixing chamber.}
        \label{fig:SiPMinstallation}
    \end{subfigure}
    \caption{Measurement setup\label{fig:meassetup}}
\end{figure}

\begin{figure*}[h]
    \centering
    \includegraphics[width=0.9\linewidth]{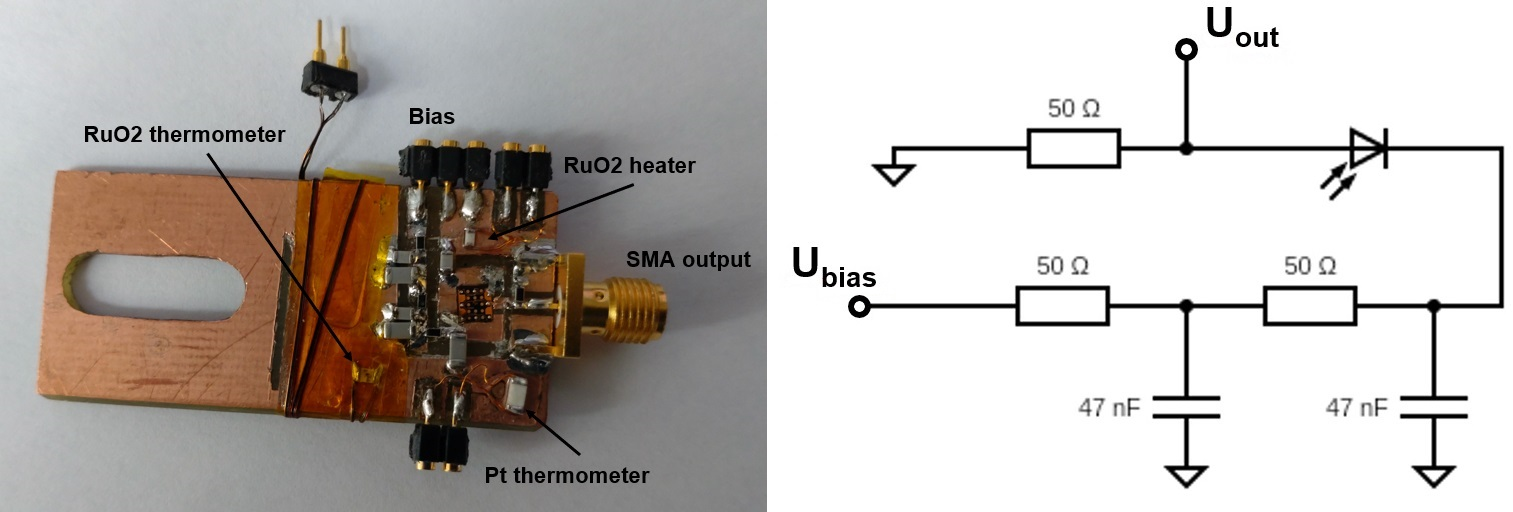}
    \caption{Picture and schematic of the basic SiPM biasing board.}
    \label{fig:biasboard}
\end{figure*}

In our measurement setup (fig.~\ref{fig:meassetup}),  each SiPM was mounted to a biasing and measurement board (fig.~\ref{fig:biasboard}) and placed on the mixing chamber of a homemade wet dilution refrigerator system. The biasing board design is based on the one used by Wiesinger \cite{OnSemi4K}. In addition to the biasing circuit itself, the bias board includes two thermometers, a platinum chip and a RuO$_2$ chip, in order to cover the entire temperature range from room temperature to below $100 \un{mK}$. There is also a second RuO$_2$ chip on the board, used as a heater for fine tuning of the SiPM temperature. The SiPM was illuminated with an LED from room temperature, coupled into the cryostat via UV-compatible optical fiber (\href{https://www.thorlabs.com/thorproduct.cfm?partnumber=UM22-400}{Thorlabs UM22-400}). 

Additionally, a rotatable shutter thermalized at the dilution fridge mixing chamber was placed between the fiber and the SiPM. The main purpose of the shutter was to prevent room temperature infrared photon leakage via the fiber during dark count measurements. The fiber has also been thermalized at several temperatures ($4\un{K}$, $1\un{K}$, mixing chamber) along the way, to minimize heat leak from the fiber connection at room temperature. The LED output was purposely left unfocused to minimize total photon flux through the fiber. The output from the fiber was also left uncollimated, resulting in a beam diameter of roughly $4 \mathrm{~to~} 5\un{mm}$ at the SiPM.

Our ability to effectively stabilize the system temperature during the cooldown and warmup processes is limited, which restricted our ability to perform measurements between liquid helium (LHe, $4.2\un{K}$) and liquid nitrogen (LN2, $77\un{K}$) temperatures, as well as between liquid nitrogen and room temperatures. Due to low thermal coupling of the SiPM to the dilution fridge mixing chamber, we were able to overheat the SiPMs and stabilize the temperature for measurements up to $\sim40\un{K}$.

Pulseform data acquisition was done via a Keysight DSOX1102G oscilloscope, and the illuminating LED was controlled via the same oscilloscope's signal generator. The SiPM bias voltages are controlled and photocurrents were measured via a Keithley 6487 current source. In our amplifier chain we used a ZFL-1000NL and a ZFL-1000NL+ at room temperature. Code and data are available on GitHub \cite{Kiilerich_SiPM_measurement_scripts}. 

For breakdown voltage ($V_{bd}$) measurements we used several techniques. At room temperature, we measured the breakdown voltage directly from the dark current as the crossing point of zero current with a linear fit on a $\sqrt{I}/V$ plot \cite{introtosipmonsemi}. At cryogenic temperatures the dark count rate is too low to generate a measurable dark current, necessitating a different approach. We used two methods to extract the breakdown voltage in this regime; firstly, we measured the breakdown voltage from the pulse height of our SiPM versus voltage \cite{OnSemi4K}. $V_{bd}$ was subsequently determined as the crossover point of pulse height with zero. The second method is to use low LED power, $10 \un{\mu A}$ current for the blue LED and $50 \un{\mu A}$ for the UV LED, to illuminate the SiPM at a low signal level in lieu of dark current. $V_{bd}$ can then be measured with the same methodology as in the room temperature measurements. The results from these different methods are in agreement with each other within the margin of error.

A more detailed description of the measurement setup and data analysis methodology is given in the reference \cite{TomGradu}.

\section{Results}

\subsection{Pulse shape}

\begin{figure}[h]
    \centering
    \begin{subfigure}{0.95\linewidth}
        \includegraphics[width=\linewidth]{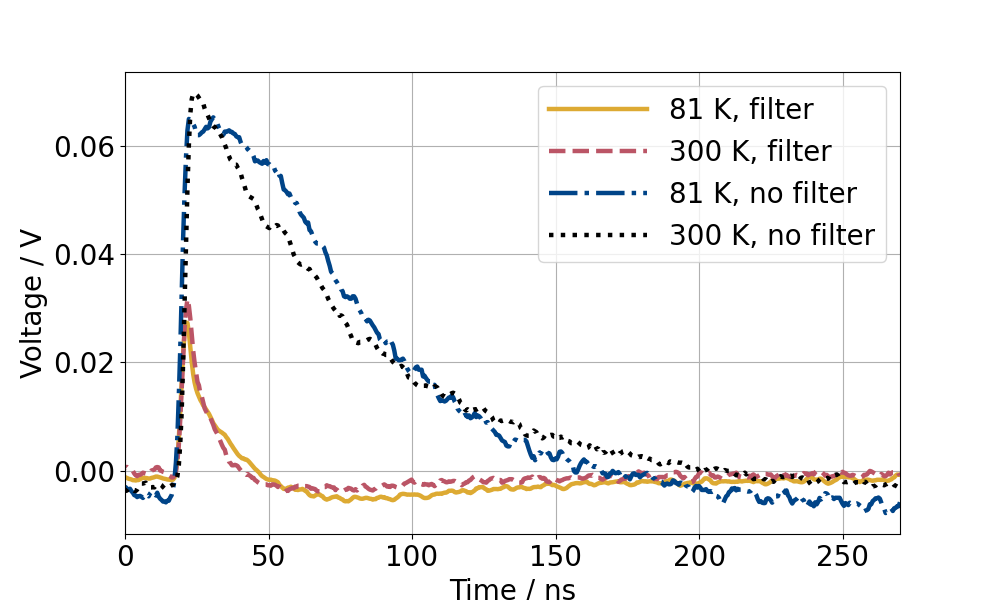}
        \subcaption{Hamamatsu\label{fig:Hamapulseshape}}
    \end{subfigure}
    \\
    \begin{subfigure}{0.95\linewidth}
        \includegraphics[width=\linewidth]{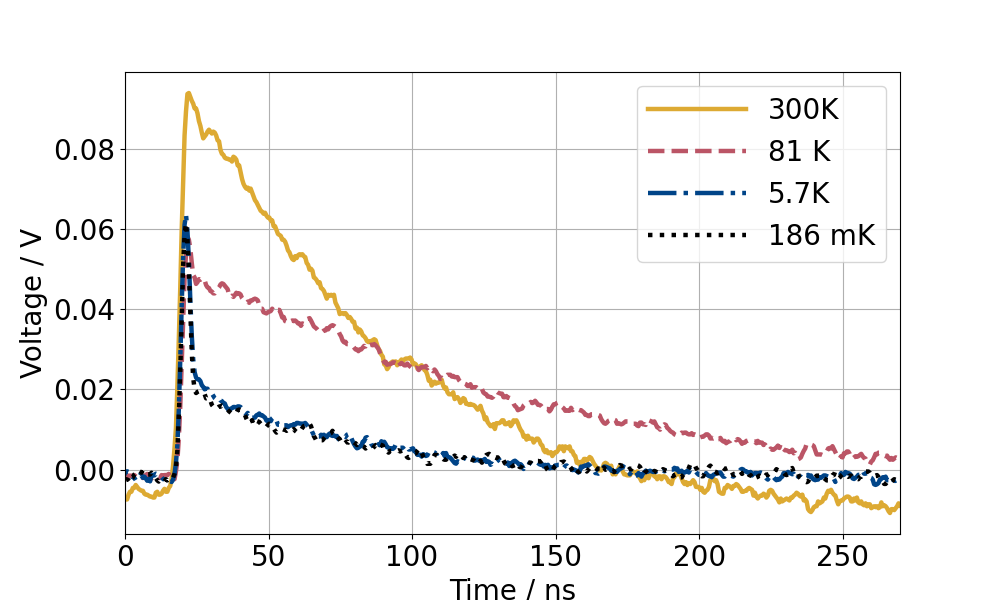}
        \subcaption{Onsemi \label{fig:Onsemipulseshape}}
    \end{subfigure}
    
    \caption{SiPM pulse shapes, measured as a 30 pulse average. Filter refers to the high-pass RC filter described in section \ref{sec:MeasSetup}.\label{fig:pulseshapes}}
\end{figure}

Similar to previous experiments \cite{OnSemi4K}, we saw pulse shape changes due to quench resistor temperature dependence when cooling down to cryogenic temperatures (fig.~\ref{fig:Onsemipulseshape}) with our Onsemi chip. At room temperature the pulse has a fast voltage peak from the avalanche, followed by a decay back to zero voltage with characteristic time around 100 ns. At lower temperatures the pulse shape is divided into two components, a fast and slow component. The fast component decays with characteristic time below 10 ns, whereas the slow component decays with characteristic times above 100 ns. Additionally, the peak pulse voltage decreased by roughly 30\% when cooled down from the room temperature to liquid nitrogen temperature. No further degradation of maximum pulse voltage was seen on cooling below this point, and below 10 K the pulse  shape did not change at all. 

For the Hamamatsu chip, we saw minor pulse shape changes from room temperature to liquid nitrogen temperature (see fig.~\ref{fig:Hamapulseshape}). From liquid nitrogen down to $150\un{mK}$, no further changes were detected. This is due to a metallic quench resistor having orders of magnitude smaller temperature dependence in the low temperature regime than that of semiconductor resistors. For Hamamatsu, we additionally utilized a high-pass RC filter to simulate a fast output mode and increase our pulse counting resolution at higher photon fluxes.

At room temperature, the Onsemi fast output channel could be used to generate extremely sharp rise time pulse signals for pulse counting purposes without the long tail characteristic of a recharging SPAD. However, the on-chip filters on the Onsemi chip did not function properly when cooled down, and below $110\un{K}$ the fast output channel created a lot of excess noise. This rendered the channel inoperable at low temperatures. One should instead use a room temperature high-pass filter on the normal output channel, as was done with the Hamamatsu chip.

The relevance of the pulse shape changes depends on the application the sensor is used for. When used as a single photon counter for low photon fluxes, the shape changes do not affect the sensor functionality in a major way, since in this type of measurement one should count the pulses by rising edge. In these measurements, a high-pass filter should regardless be implemented to cut the slow signal tail for a better SNR. When used for measuring large photon fluxes where the signal will be calculated as an integral of the signal, one must take into account the signal shape, as it strongly affects the integral of the pulse shape and the detected photo-current.

\subsection{Dark count rate}
An essential feature of the measurements at low temperatures is that a true dark condition can be realized when excitation via light or thermal effects are fully frozen out. Starting from room temperature, we observed a fast drop of the dark count rate (DCR), roughly by the factor of 2 for each 10 K, consistent with the manufacturers' specifications. Already at $\sim$ 100 K we noticed that the dark count rate strongly depends on the position of the shutter above SiPM. Closing the shutter decreased the DCR by more than two orders of magnitude, reaching dark count rates well below 1 Hz. With the closed shutter we have a true dark condition; the upper end of the optical fiber is located in a room temperature light tight box, but there can remain a tiny flux of thermal IR photons leaking into the low temperature part of the apparatus. These photons have sufficient energy to create avalanches and generate non-zero photocurrent. This leakage can be blocked with a cold shutter after the fiber.

Once the chips were cooled below 100 K and the shutter was closed, we could only see occasional pulses at count rates below 1 Hz. This is expected, since at temperatures below $200\un{K}$, the dark count rate is dominated by band-to-band tunneling \cite{studiesofsipmatcryo}, which has a significantly lower excitation rate compared to thermally excited dark counts at higher temperatures. No further reduction of the DCR was observed on cooling into sub-Kelvin range. We suggest that the residual DCR is caused by scattered thermal radiation photons which leak inside the vacuum can of our refrigerator via imperfect radiation shields.  A tiny fraction of them can reach the chip through the gap between the shutter plate and the chip surface. Another possibility could be e.g. cosmic rays or high energy elementary particles, which penetrate through the metal enclosures of our cryostat. We did not perform a detailed quantitative study of DCRs below 1 Hz.

Closing the shutter to achieve a true dark condition was a key factor for observing the second breakdown and correct determination of the operating voltage range, as described in the next section.

\subsection{Breakdown voltages and tunneling effect}
The breakdown effect of a SiPM is seen as a rapid increase of photosensitivity of the detector above a certain bias voltage, the breakdown voltage $V_{bd}$. At this point, the electric field is high enough that a charge carrier generated by an incident photon will gain sufficient kinetic energy over the mean free range in the bulk to generate another charge-carrier pair upon impact, leading to an avalanche. At room temperature, the charge carrier may be created by thermal excitations, without illumination by light. This leads to the fairly high dark count rate, increasing proportionally to the overvoltage above $V_{bd}$. The breakdown voltage is typically temperature dependent, due to increased carrier mobility at low temperatures \cite{lowtempsilicon}.

At sufficiently low temperatures (below $\sim 100\un{K}$), the thermal excitations are frozen out, and if no photons are coming to the SiPM, thermally excited breakdowns do not spontaneously occur even at fairly large overvoltages. Therefore, under true dark conditions we need to send a small flux of photons, which allows measurements of the $V_{bd}$, as explained in Section \ref{sec:MeasSetup}.

\begin{figure}
    \centering
    \includegraphics[width=\linewidth]{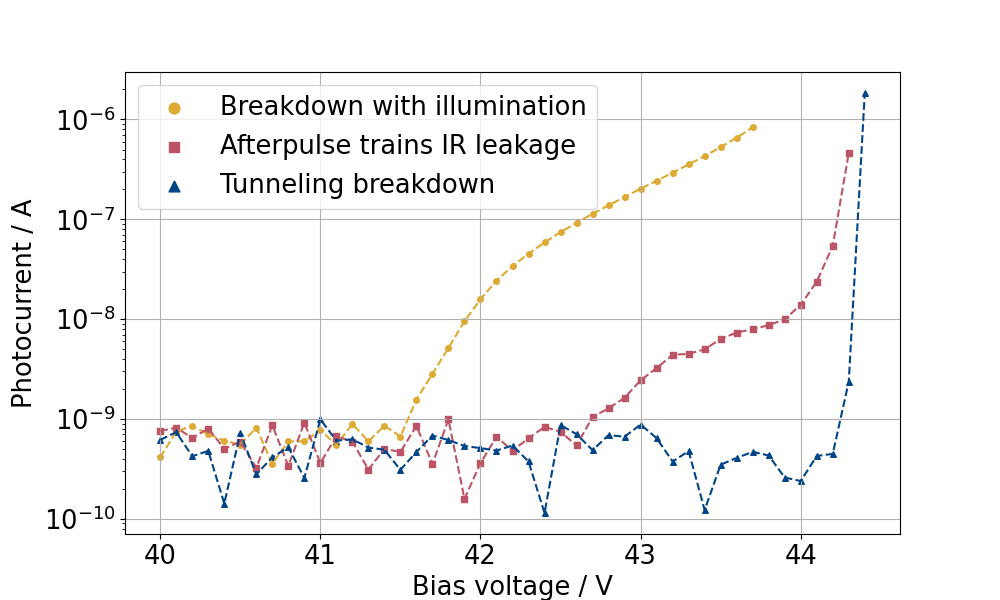}
    \caption{Different types of breakdown behaviour on the Hamamatsu chip, measured at $1 \un{K}$. Normal breakdown occurs at $V_{bd}\approx41.5 V$, tunnelling breakdown is seen at $V_{t}\approx44.2 V$.}
    \label{fig:breakdowns}
\end{figure}

When increasing the overvoltage further in dark conditions at temperatures below $40\un{K}$, we found that the photocurrent exhibited a sudden and steep growth indicating the appearance of a second breakdown at a voltage $V_t$, which depends strongly on temperature. For the Onsemi SiPMs we found that the difference between these two breakdown voltages exceeds $12\un{V}$ at and above $40\un{K}$, which would indicate a cryogenic operational voltage range substantially larger than the $5\un{V}$ overvoltage recommended by the manufacturer for room temperature applications. We demonstrate the typical behaviour of both breakdown phenomena in fig.~\ref{fig:breakdowns}.

We suggest that the second breakdown could be caused by a tunneling phenomenon, when the bias voltage is high enough for charge carriers to directly tunnel from the valence band (band-to-band tunneling), or intermediate impurity-induced energy states to the conduction band (trap-assisted tunneling).

This effect has not been previously reported and the physical basis for the second breakdown has not previously been explained. In their characterization experiments,  Zhang et al \cite{Zhang_SiPMs} reported a rapid rise in photocurrent at higher overvoltages, which they claimed to be due to increased afterpulsing probability causing self-sustaining afterpulse trains. We were able to reproduce this behaviour (orange line in fig.~\ref{fig:breakdowns}) with an open shutter. We deduce that by blocking IR leakage from room temperature with our cryogenic shutter, we create a true dark condition, necessary for observation of the tunneling breakdown.  

\begin{figure*}
    \centering
    \begin{subfigure}{0.47\textwidth}
        \includegraphics[width=\linewidth]{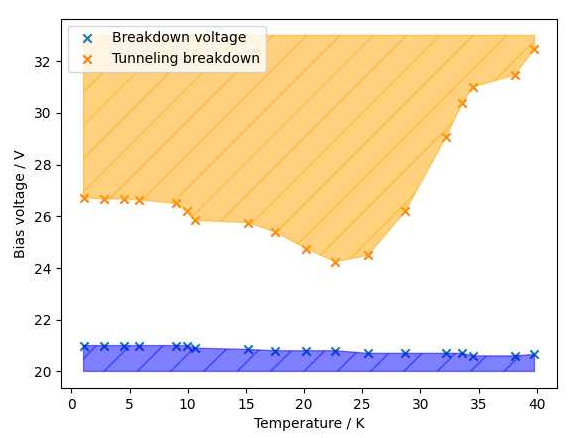}
        \subcaption{Onsemi\label{fig:OnsemiOpV}}
    \end{subfigure}
    ~
    \begin{subfigure}{0.512\textwidth}
        \includegraphics[width=\linewidth]{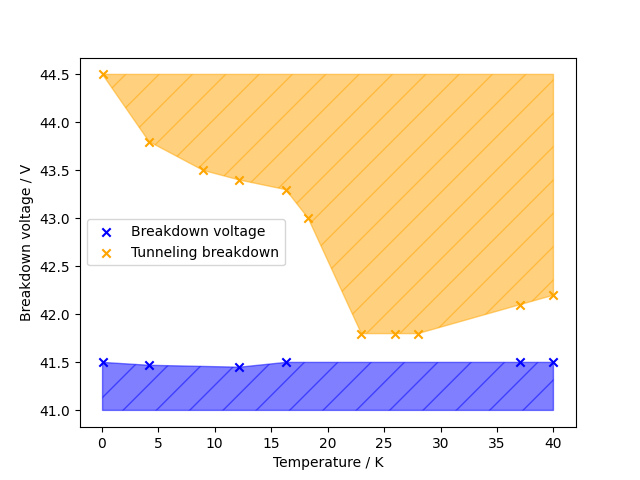}
        \subcaption{Hamamatsu \label{fig:HamamatsuOpV}}
    \end{subfigure}
    
    \caption{Operational voltage ranges at temperatures $T \leq 40\un{K}$ \label{fig:OpV}}
\end{figure*}

\subsection{Operational voltage range}
After understanding the breakdown phenomena at cryogenic and true dark conditions, we may now define an operating range for the SiPMs as the voltage range between $V_{bd}$ and $V_t$. We characterize the operating range at temperatures below $40 \un{K}$, where the second breakdown can be clearly identified. The temperature range between 77 K and 40 K is problematic for reliable stabilization of the SiPM temperature inside the vacuum can of the dilution refrigerator. As is seen in fig.~\ref{fig:OpV}, $V_t$ showed a pronounced minimum for both chips at around $23\un{K}$. The first breakdown voltage $V_{bd}$ monotonously decreased during cooldown from its room temperature value, by approximately 5 V for the Onsemi chip, and approximately 10 V for the Hamamatsu chip, reaching its minimum around LN2 temperature. $V_{bd}$ remained nearly constant throughout our primary measurement range below 40 K.

Operating the Hamamatsu SiPM near $25\un{K}$ becomes problematic, since overvoltages above $0.3\un{V}$ cannot be used. At the lowest temperatures below $5\un{K}$ the operating range increases to $\sim$ 3 V, which is comparable with the values recommended by the manufacturer at room temperature. The Onsemi device operating range increases to $> 5\un{V}$ below $5\un{K}$ and does not change at lower temperatures. A remarkably large operating range of more than $12 \un{V}$ is observed for the Onsemi chip at $\sim 40\un{K}$. The overall behaviour of both chips is similar to reports on earlier experiments on the Hamamatsu chip by Zhang et al \cite{Zhang_SiPMs}.

\subsection{Afterpulsing}

Afterpulsing behavior is typically seen as an appearance of secondary pulses followed the main pulse with a delay comparable with the pulse relaxation time.  An afterpulse is a secondary avalanche of a SPAD typically caused by charge carriers from the initial avalanche getting trapped in impurities of the silicon \cite{AfterPulseImpurities}, and causing a second avalanche when released. A single primary event may have several afterpulses, either due to multiple trapped charge carriers from the primary pulse released at different intervals or by having an afterpulse trigger a second afterpulse. 

Afterpulses are easiest to detect by observing dark counts of the device, since due to the low dark count rate of around $1\un{Hz}$ at cryogenic conditions, the likelihood of detecting multiple dark counts in a short time window is significantly lower than the afterpulsing probability.

The afterpulsing effects essentially function as an additional noise source for both integrating measurements for high photon fluxes and pulse-counting measurements for low photon fluxes. Both the long and short delay afterpulses distort the integral of a single detection event signal, and will therefore distort any averaging measurements. The long delay pulses additionally create additional pulse counts, indistinguishable from primary detection events, creating a systematic error in a pulse count measurement. The short delay pulses do not cause this issue, as they can be easily distinguished by a significantly lower pulse voltage.

Afterpulsing creates a systematic error, which always overestimates the real photon flow. For high photon flux measurements where we are detecting a photocurrent, afterpulsing effectively functions as an extra gain factor for the signal. For experiments where measuring the absolute photon flux is important, the afterpulsing effects need to be characterized and accounted for. However, if one is only interested in the relative photon flux determination, the afterpulses should have negligible effect on measurement error.

At temperatures below 40 K, we detected a significant increase in the probability of afterpulsing and an increase in the average time delay between the primary pulse and afterpulse(s). Together, these effects occasionally combine to create long-lasting afterpulse trains, where successive afterpulses cause further afterpulses. This cascade of afterpulses can last for a long time at cold temperatures, up to roughly $1\un{ms}$. As the afterpulsing probability also increases with overvoltage, these afterpulse trains can limit the functionality of the SiPMs at low temperatures. This is due to increased noise characteristics and heating effects from afterpulse-train induced photocurrent, as previously reported \cite{Zhang_SiPMs}. 

We did not do a quantitative analysis of the afterpulsing effects in the scope of this study, as the effect has low relevance to our intended application of the SiPM chip as a fluorescence detector.

\begin{figure}
    \centering
    \begin{subfigure}{0.4935\linewidth}
        \includegraphics[width=\linewidth]{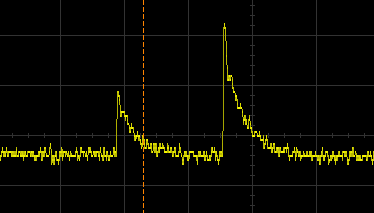}
        \subcaption{\label{fig:afterpulselong}}
    \end{subfigure}
    ~
    \begin{subfigure}{0.465\linewidth}
        \includegraphics[width=\linewidth]{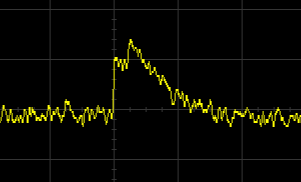}
        \subcaption{\label{fig:afterpulseshort}}
    \end{subfigure}
    \\
    \begin{subfigure}{0.7\linewidth}
        \includegraphics[width=\linewidth]{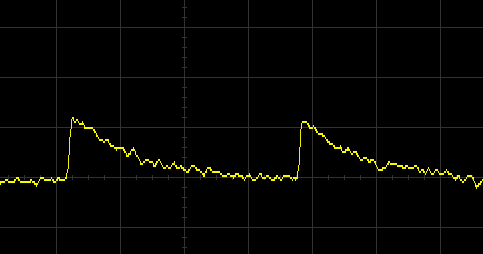}
        \subcaption{\label{fig:afterpulsenocross}}
    \end{subfigure}
    
    \caption{Afterpulsing pulse shapes, measured from the Hamamatsu chip. a) Afterpulse with a long delay and crosstalk on the secondary pulse b) Afterpulse with a short delay c) Afterpulse with no crosstalk \label{fig:afterpulses}}
\end{figure}

\subsection{Photon detection efficiency}

For the measurement of photon detection efficiency (PDE) at cryogenic temperatures, the main difficulty is having a calibrated photon flux to calculate the detection efficiency. As we do not have access to a calibrated cryogenic photon source, we instead opted for a relative measurement of PDE.

The room temperature specifications of the sensors are listed in table \ref{tab:RTspecs}. Using these data as a calibration point for our light source, we are able to calculate the relative PDE of the sensor at lower temperatures without knowing the absolute value of the incident photon flux.

\begin{table}
    \renewcommand{\arraystretch}{1.2}
    \centering
    \begin{tabular}{c|c|c}
         & Hamamatsu & Onsemi \\\hline\hline
         $V_{\mathrm{bd}}$ [V] & $51.3$ & $24.5$ \\\hline
         PDE(470 nm) [\%] & 28 & 31 \\\hline
         PDE(275 nm) [\%] & 17 & 7.5
    \end{tabular}
    \caption{Room temperature specifications for our SiPMs \cite{hamamatsudatasheet,onsemidatasheet}. Breakdown voltages $V_{bd}$ measured by us, due to high error margins in manufacturer specs.\label{tab:RTspecs}}
\end{table}

To measure the relative PDE, we pulse the LED repeatedly with constant pulse parameters, we can determine from the statistics the mean number of detected events via a Poissonian fit. From this we subtract the dark count rate to determine the mean number of photons per pulse \cite{TomGradu,OnSemi4K}. We then use the manufacturer specifications for the room temperature PDE values to calibrate our absolute PDE value relative to the photon flux from our LED, and use this as the comparison point for our low temperature measurements. In order to avoid afterpulsing effects, we chose a low overvoltage of $2.5\un{V}$ for the PDE measurements.

\begin{figure*}
    \centering
    \includegraphics[width=0.98\textwidth]{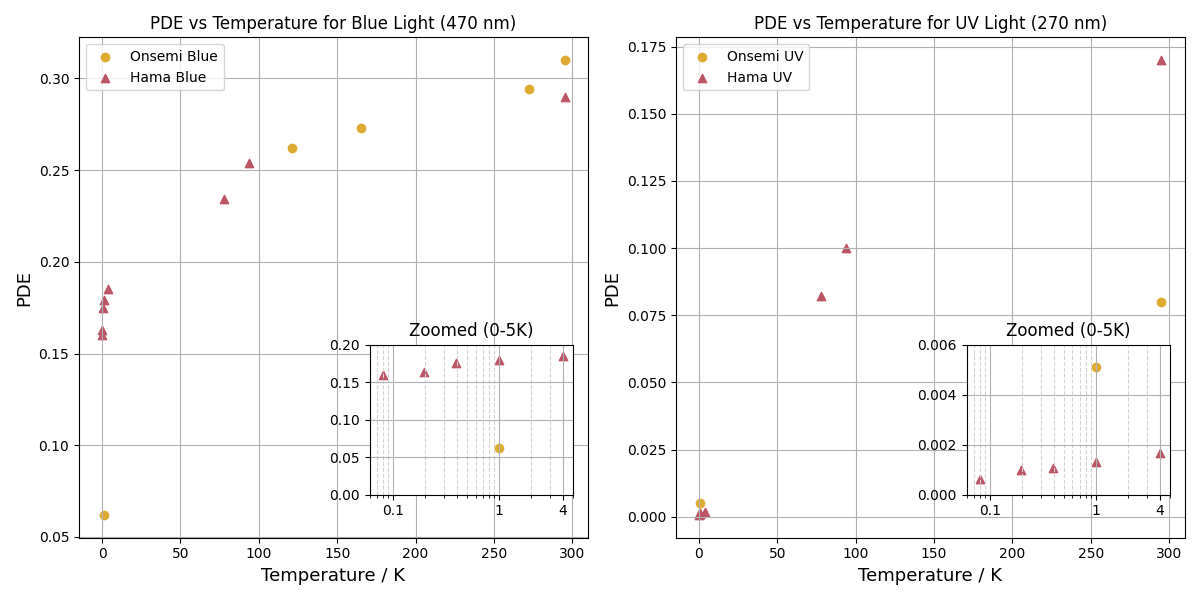}
    \caption{PDE as a function of temperature, with zoom window on the low temperature regime. Each point measured at a constant $2.5 \un{V}$ overvoltage.\label{fig:PDE}}
\end{figure*}

When cooled down from room temperature to temperatures below $1\un{K}$, we see a large drop in PDE (fig.~\ref{fig:PDE}). For both tested sensors, the drop in PDE is significantly larger for the UV than for the visible blue light. Our measurement data is in good agreement with previous measurements for both devices \cite{Zhang_SiPMs,HamamatsuCryo2,OnSemi4K}.

Both SiPMs are significantly more sensitive to visible blue light than to UV light at cryogenic temperatures, with a difference of 1 to 2 orders of magnitude. Therefore for detection of scattered or fluorescent UV light, a measurement scheme with frequency converters based on TPB \cite{CryoTPB} would remain significantly more efficient even after conversion losses.

\subsection{Thermal cycling tolerance}

We monitored the thermal cycling tolerance of the chips in order to get an idea of their useful lifetime in cryogenic experiments. Through several thermal cycles, we saw no notable drop in PDE or other performance characteristics, albeit with the Onsemi chip, which has a protective glass layer covering the SiPM pixels, we saw damage in the form of cracks on the glass layer. From our PDE data we saw no performance impact from these cracks when comparing room temperature PDE before and after thermal cycling.

The cracks in the glass cover are typically formed during the first cooldown, and there is no notable progression in the damage over subsequent cooldowns. Due to this and the aforementioned good retention of actual performance characteristics, the Onsemi SiPMs appear to be well tolerant to thermal cycling, and can be operated over several thermal cycles without replacement.

The Hamamatsu chips we tested have no protective glass cover, and therefore no cracking or other damage related to thermal cycling was observed.

\begin{figure}[h]
    \centering
    \begin{subfigure}{0.48\linewidth}
        \includegraphics[width=\linewidth]{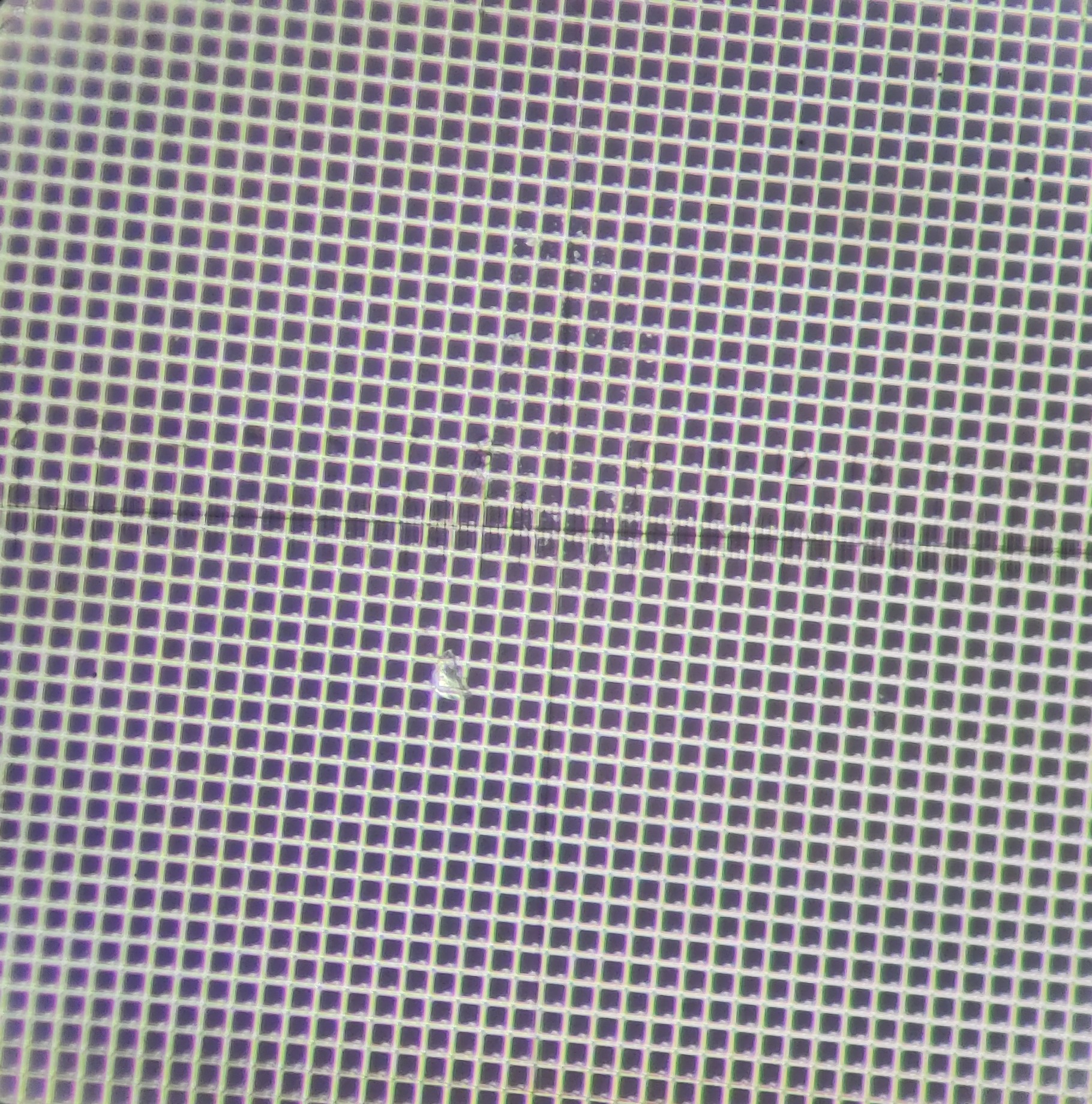}
        \caption{Hamamatsu, microcell dimensions $50\times 50 \un{\mu m^2}$\label{fig:GlassCrackA}}
    \end{subfigure}
    ~
    \begin{subfigure}{0.48\linewidth}
        \includegraphics[width=\linewidth]{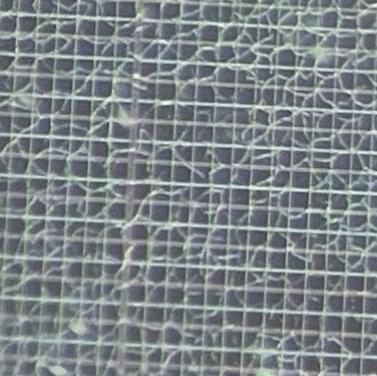}
        \caption{Onsemi, microcell dimensions $40\times 40\un{\mu m^2}$\label{fig:GlassCrackB}}
    \end{subfigure}
    \caption{Thermal cycling surface damage. a) Hamamatsu without glass cover b) Onsemi with glass cover \& cracks\label{fig:GlassCrack}}
\end{figure}

\section{Conclusions}

We have characterized the functionality of two commercial SiPM models as low-flux photon detectors in a wide temperature range from room temperature down to 90 mK. We found that while their performance drops with temperature, they are functional with a PDE above $10\un{\%}$ throughout the entire temperature range in the visible blue light range, and down to roughly $100\un{K}$ at UV wavelengths. Hamamatsu S13371-6050CN showed somewhat better PDE than Onsemi, although with a smaller operating voltage range. The operating voltage for Hamamatsu detectors is nearly a factor of 2 larger than for Onsemi detectors, which results in a larger Joule heating for the same photo-current. This can become an issue at low temperatures. 

The relaxation after an avalanche became very slow for Onsemi detectors at low temperature due to semiconductor quench resistors. However, well resolved and sharp signal peaks are observed in the beginning of an avalanche pulse, which can easily be detected and counted. Metal based quench resistors utilized in Hamamatsu detectors therefore do not provide a significant benefit at low temperature.

For the purpose of low temperature fluorescence detection in the UV range, we suggest using a wavelength shifting coverage for conversion of the UV photons to the blue light. 

For follow-up studies we would recommend a more thorough analysis of the afterpulsing behaviour to confirm the assumption of signal linearity, and to more accurately characterize the exact afterpulse probability effect. Building a better theoretical understanding of the proposed tunneling breakdown effect might provide information on how to optimize SiPM design for maximum gain at cryogenic temperatures. Extensive testing with large sample sizes of detectors from different production batches and over a larger number of cycles would also provide worthwhile information on the reproducibility of cryogenic SiPM characteristics.

\section*{Statements and declarations}
\subsection*{Acknowledgments}
This project was supported by the Jenny and Antti Wihuri foundation.

\subsection*{Funding}

No funds or grants were received for the purpose of this work. Dilution fridge operation costs were covered by the Wihuri Physical Laboratory, with funding received from the Jenny and Antti Wihuri foundation.

\subsection*{Competing interests}

All authors certify that they have no affiliations with or involvement in any organization or entity with any financial interest or non-financial interest in the subject matter or materials discussed in this manuscript.

\subsection*{Author contributions}
All authors contributed to the concept and the design of the experimental setup. Setting up and running the experiment were performed by Otto Hanski, Tom Kiilerich and Slava Dvornichenko. Data collection and was done by Tom Kiilerich, data processing and analysis was done by Tom Kiilerich and Otto Hanski. The first draft of the manuscript was written by Otto Hanski and Tom Kiilerich. All authors commented on previous versions of the manuscript and approved its final version. 

\subsection*{Data Availability Statement}
 The datasets generated and/or analyzed during the current study are available from the corresponding author on request. Unsorted data is also available in the related data collection uploaded in Github \cite{Kiilerich_SiPM_measurement_scripts}.


\bibliography{sn-bibliography}% common bib file

\end{document}